\documentclass[12pt,a4paper]{iopart}
\usepackage{iopams}
\usepackage{amsopn}
\usepackage{mathrsfs}
\usepackage{latexsym}
\usepackage{citesort}
\usepackage{bm}
\usepackage{tensor}

\usepackage{graphicx}
\usepackage{caption}
\usepackage{amsopn}
\DeclareMathOperator{\sech}{sech}

\def\bra#1{ \left\langle #1 \right|}
\def\ket#1{\left| #1 \right\rangle }

\providecommand{\abs}[1]{\bigl| #1 \bigr|}

\def\vec#1{\rm{\textbf#1}}

\begin{document}

\title{Effective thermodynamics of isolated entangled squeezed and coherent states}

\author{King Karl R Seroje, Rafael S {dela Rosa}, Francis N~C~Paraan\footnote[1]{To whom correspondence should be addressed}}
\address{National Institute of Physics, University of the Philippines Diliman, 1101 Quezon City, Philippines}
\ead{\mailto{kingkarlseroje@nip.upd.edu.ph}, \mailto{rdelarosa@nip.upd.edu.ph} and \mailto{fparaan@nip.upd.edu.ph}}
\date{\today}

\begin{abstract}
The R\'enyi entanglement entropy is calculated exactly for mode-partitioned isolated systems such as the two-mode squeezed state and the multi-mode Silbey--Harris polaron ansatz state. Effective thermodynamic descriptions of the correlated partitions are constructed to present quantum information theory concepts in the language of thermodynamics. Boltzmann weights are obtained from the entanglement spectrum by deriving the exact relationship between an effective temperature and the physical entanglement parameters. The partition function of the resulting effective thermal theory can be obtained directly from the single-copy entanglement.
\end{abstract}




\section{Introduction}\label{sec:intro}

Entanglement is a central topic in quantum information theory. It is viewed as a resource for manipulating quantum information and accomplishing quantum computing tasks. In an isolated pure state $\rho_{AB} = \ket{\psi}\bra{\psi}$, entanglement is manifested as quantum correlations between its partitions $A$ and $B$. That is, the results of measurements done on a partition, say $A$, are statistically distributed according to the reduced density operator $\rho_A = \tr_B \rho_{AB}$, where the partial trace $\tr_B(\cdot)$ is a quantum average done over a complete set of states for the complementary partition $B$. For these bipartite pure states, $\rho_A$ and $\rho_B$ have identical spectra (eigenvalues) and hence entanglement measures can be calculated from either reduced density operator. When the partitions $A$ and $B$ are \textit{unentangled} both $\rho_A$ and $\rho_B$ are projectors onto a single state vector. On the other hand, when $A$ and $B$ are \textit{maximally entangled} $\rho_A$ and $\rho_B$ correspond to uniform distributions over their respective eigenvectors. That is, there is more entanglement in $\rho_{AB}$ when there is less certainty in the state of a given partition when the effects of its complement are averaged over. A similar relationship between the uncertainty in knowing the (reduced) state of a subsystem and the strength of its coupling with its complement or environment is well-established in statistical mechanics \cite{feynman1963a}.

The main pedagogical objective in this article is the presentation of several key concepts of quantum information theory using the language of statistical mechanics and thermodynamics. The construction of effective thermal models for a partition is a very useful concept that underlies more advanced ideas such as the emergence of area laws in the scaling of entanglement entropy \cite{srednicki1993a,wolf2008a,eisert2010a} and the replica trick of quantum field theory \cite{ryu2006b,calabrese2009a}. The systems under consideration here are also particularly illustrative: For example, uncertainty and fluctuations in a mode-partitioned squeezed state are identical to that of a harmonic oscillator at thermal equilibrium. Some of the results given in this paper have been demonstrated in the past \cite{han1999a,botero2003a}, but here we emphasize the consonance of the quantum information and statistical mechanics perspectives. Additionally, we also present some new exact results such as the detailed calculation of the R\'enyi entanglement entropy in bosonic two-mode squeezed states and a multi-mode Silbey--Harris state \cite{silbey1984a}, and the identification of the single-copy entanglement \cite{vidal1999a} as a free energy analogue \cite{peschel2005} for these systems. For a broader overview of the research on entanglement in many-particle systems, the reader may refer to extensive technical reviews found in the literature \cite{amico2008a,horodecki2009a} as well as others at a more introductory level \cite{peschel2012a}. 

The mode-entangled examples studied here involve coherent (displaced) and squeezed boson modes that are encountered in, for example, quantum optical systems \cite{braunstein2005a,eichler2011a}, ultracold coherent matter systems \cite{parkins2006a,esteve2008a}, and nanomechanical systems \cite{xue2007a}. We partition these composite states with respect to the mode indices $A$ and $B$ and analytically obtain the reduced density operator for a partition $\rho_A$. This allows us to calculate entanglement measures (\sref{sec:entanglement}) for these bipartite pure states, such as the R\'enyi entanglement entropy \cite{vidal2000a} and entanglement spectrum \cite{li2008a}. The entanglement spectrum is further used to construct effective thermal models that yield identical fluctuations in observables that are local to a given partition (mode-local). We first consider several cases of two-mode coherent and squeezed states (\sref{sec:resultsCS}) and then proceed to the case of the multi-mode Silbey--Harris state (\sref{sec:resultsSH}).

\section{Quantifying entanglement}\label{sec:entanglement}

The reduced density operator $\rho_A$ is easily obtained when the composite state vector $\ket{\psi}_{AB}$ is normalized and expressed in Schmidt decomposed form \cite{nielsenchuang}:
\begin{equation}\label{schmidt}  
 \ket{\psi}_{AB} = \sum_{n=1}^d \lambda_n \ket{n}_A\otimes\ket{n}_B , \quad {\rm{with\quad }}\bra{m}_{x\,}\ket{n}_y= \delta_{mn}\delta_{xy}, 
\end{equation}
and $\ket{n}_x$ is a Schmidt state vector with a subscript $x\in\{ A,B\}$ that labels the mode partition, $\lambda_n \ne 0$ is the corresponding Schmidt coefficient, and $d$ is the Schmidt number or Schmidt rank of $\ket{\psi}_{AB}$. The Schmidt coefficients satisfy the normalization condition $\sum_{n=1}^{d} \abs{\lambda_n}^2 = 1$.
For brevity, we will omit the mode indices for the state vectors $\ket{n}_A = \ket{n}$ and the tensor product operator $\otimes$ when context is sufficient. Evaluating the partial trace over mode $B$ then yields a diagonal representation for $\rho_A$ in the orthonormal basis of mode $A$ Schmidt vectors:
\begin{equation}
	\rho_A = \tr_B\ket{\psi}_{AB}\bra{\psi}_{AB} = \sum_{n=1}^d p_n \ket{n}_A\bra{n}_A,
\end{equation}
which is an operator of rank $d$. The eigenvalues of this reduced density operator are the complex squares of the Schmidt coefficients $p_n \equiv \abs{\lambda_n}^2$. The state $\ket{\psi}_{AB}$ is separable (unentangled) with respect to the bipartition $A\cup B$ if and only if there is exactly one non-zero Schmidt coefficient (or, equivalently, the Schmidt rank is $d=1$). Furthermore, if there is more than one non-zero Schmidt coefficient ($d>1$) then the state is entangled. For the special case in which $\abs{\lambda_n}^2 = 1/d$ for all $n \in \{1,2,\dots,d\}$, the state is maximally entangled. Quantitative measures of entanglement in the bipartite pure state $\ket{\psi}_{AB}\bra{\psi}_{AB}$, such as the entanglement spectrum and entanglement entropy, are discussed in more detail below.

\subsection{Entanglement spectrum}

The eigenvalues of the reduced density operator $\rho_A$ constitute a probability distribution over the Schmidt state projectors of the reduced system since $\sum_{n=1}^d \abs{\lambda_n}^2 =1$. This set of eigenvalues is referred to as the entanglement spectrum \cite{li2008a} and may be used to construct an effective thermodynamic (canonical) ensemble over the Schmidt vectors of $A$ through 
\begin{equation}
	\rho_A = \frac{\exp \bigl({-\beta \sum_n E_n \ket{n}\bra{n} \bigr)}}{Z},\qquad Z = \tr \exp\biggl({-\beta \sum_n E_n \ket{n}\bra{n}}\biggr).
\end{equation}
This identification allows one to define an effective reciprocal temperature $\beta$ and a so-called entanglement Hamiltonian $H = \sum_n E_n \ket{n}\bra{n}$. We emphasize that this is an effective thermodynamic analogy, since the composite pure state has no defined temperature (it is isolated). However, such effective approaches have been used to investigate the mechanisms behind entanglement generation, area laws, and some critical phenomena \cite{li2008a,peschel2011a,dechiara2012a,wong2013a,chandran2014a,schliemann2014a}. 

\subsection{Entanglement entropy}

Commonly used measures for entanglement between partitions are the von Neumann and  R\'enyi entanglement entropies, which quantify the degree of uncertainty associated with a density operator. When calculated for a reduced density operator, the R\'enyi entropy is a scalar function of the entanglement spectrum
\begin{equation}
 S_{\mu} = \frac{1}{1-\mu} \ln \sum_{n=1}^{d} p_n^\mu, 
\end{equation}
where $\mu\ge 0$ is the R\'enyi parameter. The von Neumann entropy $S_1$ is the special case $S_1 = -\sum_n p_n \ln p_n$, which is mathematically identical to the Gibbs entropy of statistical mechanics and the Shannon entropy of classical information theory. Other limiting values of the R\'enyi entropy include (1) the logarithm of the Schmidt rank $S_{\mu\to 0} = \ln d$, (2) the R\'enyi-2 entropy $S_{\mu\to 2}$, which is related to the purity $\gamma = \sum_n p_n^2 = \rme^{-S_2}$ \cite{adesso2005a,adesso2012a}, and (3) the single-copy entanglement $S_{\mu\to\infty}$ (SCE), which quantifies the maximum amount of entanglement that can be distilled from one copy of a state using partition-local operations and classical communication \cite{vidal1999a,jonathan1999a,eisert2005a,hadley2008a}. 

\section{Coherent and squeezed states}\label{sec:resultsCS}
 
We now turn our attention to the calculation of the preceding entanglement measures for several mode entangled coherent and squeezed states. Emphasis is made on the effective thermal theories for the reduced systems.

\subsection{Two-mode coherent state}
A two-mode coherent state $\ket{\alpha,\beta}^{\rm{c}}$ is obtained by applying a two-mode displacement operator $\mathcal{D}(\alpha,\beta)$ on the two-mode vacuum state $\ket{0}_A\ket{0}_B$:
\begin{equation}
	\ket{\alpha,\beta}^{\rm{c}} = \mathcal{D}(\alpha,\beta)\ket{0}_A \ket{0}_B = \rme^{\alpha a^\dagger - \alpha^\ast a+ \beta b^\dagger -\beta^\ast b} \ket{0}_A \ket{0}_B, 
\end{equation}
where the superscript ``c'' on $\ket{\alpha,\beta}^{\rm{c}}$ denotes ``coherent.'' The displacement operator can be written as a product of mode-local operators by the Baker--Campbell--Hausdorff (BCH) formula, which immediately gives the one-term Schmidt decomposition
\begin{equation}
	\ket{\alpha,\beta}^{\rm{c}} = \rme^{\alpha a^\dagger - \alpha^* a}\ket{0}_A  \otimes \rme^{\beta b^\dagger -\beta^\ast b}\ket{0}_B = \ket{\alpha}_A  \ket{\beta}_B,
\end{equation}
where a single-mode coherent state is 
\begin{equation}
	\ket{\alpha} = \rme^{-\left|\alpha\right|^2/2}\sum_{n=0}^\infty\frac{\alpha^n}{\sqrt{n!}} \ket{n}.
\end{equation}
A similar one-term result is also obtained when an arbitrary number state is displaced, yielding a coherent number state $\mathcal{D}(\alpha,\beta)\ket{m}_A\ket{n}_B$. Thus, a two-mode coherent state has zero mode entanglement entropy and, like the full composite system, each mode partition is in a pure state. This example illustrates an important property of bipartite entanglement: it is unchanged by the action of partition-local unitary operations \cite{vidal2000a}.

\subsection{Two-mode squeezed vacuum state}
\begin{figure}[t]
\begin{center}
\includegraphics[width=0.5\linewidth]{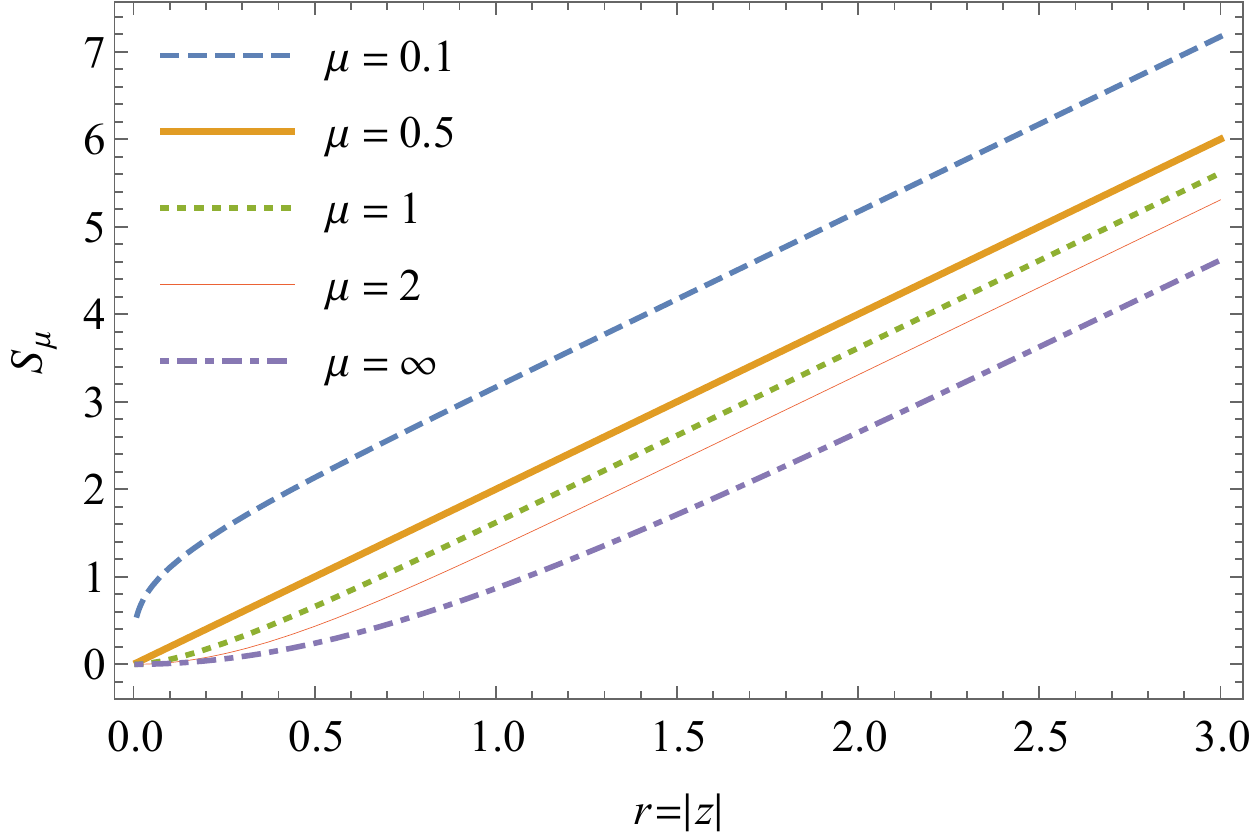}
\end{center}
\caption{\label{renyisqueezed}The R\'enyi entanglement entropy $S_\mu$ for a mode-partitioned two-mode squeezed vacuum state increases linearly with squeezing magnitude $r$ for large $r$. The R\'enyi entropy is the same for displaced squeezed states and squeezed coherent states with identical squeezing parameters.}
\end{figure}

A two-mode squeezed vacuum state $\ket{z}^{\rm{s}}$ (where the superscript ``s'' means ``squeezed'') is produced by applying the squeezing operator $\mathcal{S}(z) = \rme^{za^\dagger b^\dagger - z^* ab}$ on the vacuum state $\ket{0}_A\ket{0}_B$:
\begin{equation}\label{squeezed state}
\mathcal{S}(z) \ket{0}_A\ket{0}_B = \sum_{n=0}^{\infty} \frac{\rme^{\rmi n \theta} \tanh^n r}{\cosh r}\, \ket{n}_A\ket{n}_B,
\end{equation}
where $z=r\rme^{\rmi \theta}$ is the squeezing parameter. This is a generalization of single-mode quadrature squeezing \cite{loudon1987a} to the case of two boson modes \cite{ekert1989a,hongyi1996a}. Comparison with the defining formula \eref{schmidt} reveals that the last equality is the Schmidt decomposition for a two-mode squeezed vacuum state $\ket{z}^{\rm{s}}$. The entanglement spectrum $\{ p_n \}$ therefore consists of the probabilities
\begin{equation}
	p_n = \frac{\tanh^{2n} r}{\cosh^2 r},
\end{equation}
which are independent of the squeezing angle $\theta$. The R\'enyi entropy is readily evaluated because the entanglement spectrum forms a geometric sequence:
\begin{eqnarray}\label{eq:renyi}
S_\mu^{\rm{s}} (r) = \frac{\ln (1 - \tanh^{2\mu} r) + 2 \mu \ln \cosh r}{\mu-1}.
\end{eqnarray}
We find that the R\'enyi entropy increases monotonically with increasing squeezing magnitude $r$, where for large squeezing magnitudes $S_\mu^{\rm{s}}(r\to\infty) \sim 2\mu r/(\mu-1)$, it increases linearly with $r$ (\fref{renyisqueezed}). Thus, if the squeezing magnitude is tunable in an experimental realization of a two-mode squeezed state, then one can, in principle, generate arbitrarily large amounts of mode entanglement entropy.

Taking the limit $\mu\to 1$ leads to the known result \cite{han1999a}, which we now identify as the von Neumann entropy:
\begin{equation}
	S_1^{\rm{s}} (r) = \cosh^2 r \ln \cosh^2 r - \sinh^2 r \ln \sinh^2 r,
\end{equation}
This result is valuable since the theoretical calculation of the total von Neumann entropy in bipartite bosonic Gaussian states reduces to the evaluation of $S_1^{\rm{s}}(r)$ for two-mode squeezed states \cite{botero2003a}. Meanwhile, the R\'enyi-2 entropy is $S_2^{\rm{s}} (r) = \ln \cosh 2 r$ so that the purity is $\gamma = \sech 2r$. This expression for  the R\'enyi-2 entropy is useful because it is an experimentally accessible quantity that is important for the measurement of entanglement in Gaussian states via homodyne detections \cite{adesso2012a}. Moreover, the single-copy entanglement is $S_{\infty}^{\rm{s}}(r) = 2 \ln \cosh r$. Since the SCE is a measure of the maximum entanglement that can be distilled from a given copy of a state, this exact result is relevant to experiments that carry out such distillations on squeezed states \cite{kurochkin2014a}. 

Let us now construct an effective thermal model for the reduced single-mode state. We equate the elements of the entanglement spectrum with a corresponding effective Boltzmann weight
\begin{equation}
	p_n = \frac{\tanh^{2n} r}{\cosh^2 r} = \frac{\rme^{-\beta E_n}}{\sum_n \rme^{-\beta E_n}}.
\end{equation}
We then define an effective reciprocal temperature $\beta$ through $\rme^{-\beta\hbar\omega/2} = \tanh r$ and obtain a harmonic oscillator entanglement Hamiltonian
\begin{equation}
	H = \sum_{n = 0}^{\infty} \hbar\omega n\,\ket{n}\bra{n},
\end{equation}
with an energy shift so that the ground state energy $E_0 = 0$ (we justify this shift in the following paragraph). In other words, the statistical properties of the reduced state is the same as that of a harmonic oscillator with frequency $\omega$ in thermal equilibrium; The complementary mode $B$ interacts with $A$ as though it were in a thermal bath at an $r$-dependent reciprocal temperature $\beta$ and vice versa. 

Since the ground state energy was chosen to be $E_0 = 0$, the largest element of the entanglement spectrum in the effective thermodynamic picture is $p_0 = 1/Z$. Thus, taking the limit $\mu \to \infty$ to get the SCE from the R\'enyi entropy gives
\begin{equation}
 S_{\infty}^{\rm{s}} (r)= 2\ln \cosh r = \ln \frac{1}{1-\rme^{-\beta\hbar\omega}} = \ln Z .
\end{equation}
Hence, the evaluation of the single-copy entanglement allows one to calculate the partition function and free energy of the effective thermodynamic model. In the case of the two-mode squeezed vacuum state, the effective free energy $F = -S_{\infty}/\beta$ is 
\begin{equation}
	F = -\frac{2}{\beta}  \ln {\cosh r} = \frac{1}{\beta} \, \ln (1-\rme^{-\beta\hbar\omega}) ,
\end{equation}
which  is identical to that of a harmonic oscillator at thermal equilibrium.

\subsection{Two-mode displaced squeezed and squeezed coherent states}
We now investigate mode entanglement in states that are both displaced and squeezed. The two-mode displacement and squeezing operators generally do not commute and so we consider these cases separately. 

A two-mode displaced squeezed vacuum state is denoted as
\begin{equation}
	\ket{\alpha,\beta,z}^{\rm{cs}} = \mathcal{D}(\alpha,\beta)\bigl[\mathcal{S}(z) \ket{0}_A\ket{0}_B\bigr].
\end{equation}
The two-mode displacement operator is a product of partition-local operators and has no effect on the entanglement entropy. The R\'enyi entropy of a displaced squeezed state is therefore the same as that of the original squeezed state prior to the application of the displacement operator:
\begin{equation}
	S_\mu^{\rm{cs}} (\alpha,\beta,z) = S_\mu^{\rm{s} }(r).
\end{equation}

Next we consider entanglement in a squeezed coherent state where the order of displacement and squeezing is interchanged from the preceding example:
\begin{equation}
	\ket{z,\alpha,\beta}^{\rm{sc}} = \mathcal{S}(z) \bigl[\mathcal{D}(\alpha,\beta)\ket{0}_A\ket{0}_B\bigr].
\end{equation}
Using the BCH formula to interchange the order of squeezing and displacement operators gives
\begin{equation}
	\mathcal{S}(z)  \mathcal{D}(\alpha,\beta) = \mathcal{D}(\alpha,z\beta^\ast) \mathcal{D}(z\alpha^\ast, \beta) \mathcal{S} (z).
\end{equation}
Thus, a squeezed coherent state is also a displaced squeezed state with the same squeezing parameter and the R\'enyi mode entropy for $\ket{z,\alpha,\beta}^{\rm{sc}}$ also reduces to that of a squeezed state:
\begin{equation}
	S_\mu^{\rm{sc}} (z,\alpha,\beta) = S_\mu^{\rm{cs}} (\alpha,\beta,z) = S_\mu^{\rm{s}} (r).
\end{equation}

\section{Multi-mode Silbey--Harris state}\label{sec:resultsSH}
The multi-mode Silbey--Harris (SH) state is
\begin{equation}\label{eq:SH}
	\ket{\vec{f}^{\rm{SH}}} \equiv \frac{1}{\sqrt{2}}\,\Bigl[\ket{\uparrow}_{\rm{q}}\otimes\ket{\vec{f}\,}_{\rm{B}} - \ket{\downarrow}_{\rm{q}}\otimes\ket{-\vec{f}\,}_{\rm{B}}\Bigr],
\end{equation}
which is a variational ansatz for a spin or qubit (q) coupled to an $N$-mode harmonic oscillator or boson bath (B) via an interaction of the form $\sigma_z\sum_k(a_k+a_k^\dagger)$ \cite{silbey1984a}. This interaction leads to entanglement that can be measured and used to locate possible quantum phase transitions \cite{chin2011a}.

\begin{figure}[t]
\begin{center}
\includegraphics[width=0.5\linewidth]{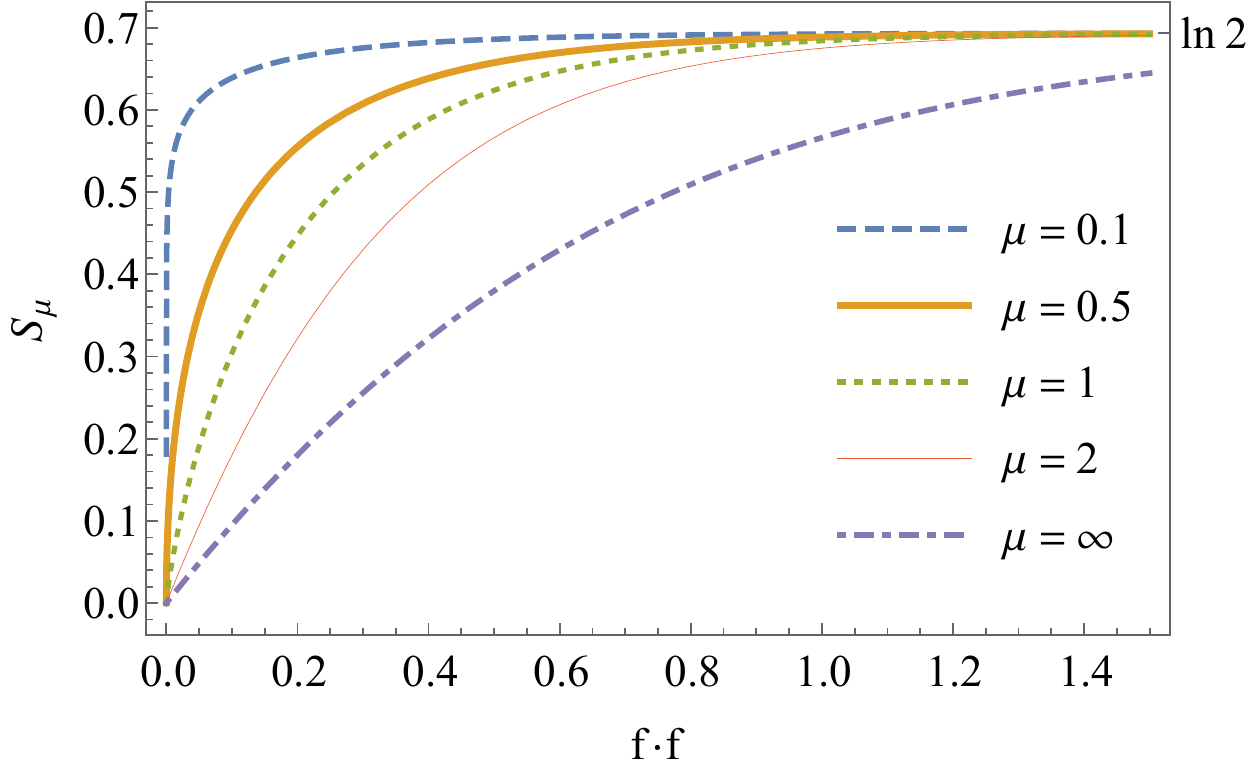}
\end{center}
\caption{\label{renyiSH}The R\'enyi entanglement entropy $S_\mu$ for a Silbey--Harris state (partitioned with respect to qubit and boson spaces) depends on all the displacement amplitudes through $\vec{f}\cdot\vec{f}$. Maximum entanglement is achieved exponentially fast with increasing $\vec{f}\cdot\vec{f}$.}
\end{figure}

The variational parameters of the SH state are the components $f_k$ of the vector $\vec{f}$ that give the displacements in an $N$-mode coherent state:
\begin{equation}
	\ket{\vec{f}\,}_{\rm{B}} = \exp\Biggl[\sum_{k=1}^{N} f_k(a_k^\dagger - a_k)\Biggr] \ket{0}_1 \ket{0}_2 \cdots \ket{0}_N	\equiv \bigotimes_{k=1}^N \ket{f_k}_k.
\end{equation}
In this work, we choose the displacements $f_k$ to be real for simplicity. The single-mode coherent states $\ket{f_k}_k$ and $\ket{-f_k}_k$ are not mutually orthogonal and so the SH state as written in \eref{eq:SH} is not Schmidt decomposed with respect to the partitioning between the qubit and boson subspaces. However, calculation of the reduced density operator on the qubit space is straightforward with the knowledge of the matrix elements (real $f$)
\begin{eqnarray}
	\bra{f}_j  \ket{f}_k = \delta_{jk}, \\
	\bra{f}_j \ket{-f}_k = \delta_{jk} \rme^{-2f^2},
\end{eqnarray}
and the fact that the trace of a tensor product of operators is equal to the product of the traces of each factor in the product. Evaluating the partial trace of $|{\vec{f}^{\rm{SH}}}\rangle\langle{\vec{f}^{\rm{SH}}}|$ over the complete set of boson number basis states yields a qubit density matrix in the $\ket{\uparrow\downarrow}$ basis
\begin{equation}
	\rho_{\rm{q}} = \frac{1}{2}
	\left( \begin{array}{cc}
1 & -c  \\
-c & 1 \end{array} \right),
\end{equation}
with the constant 
\begin{equation}
	c = \prod_{k=1}^N \bra{f_k}_k \ket{-f_k}_k= \exp\bigl(-2 \vec{f} \cdot \vec{f}\bigr).
\end{equation}
Therefore, the entanglement spectrum for the mode-partitioned Silbey--Harris state is $\{\frac{1}{2} \pm \frac{c}{2}\}$ and the R\'enyi and von Neumann entropies are
\begin{eqnarray}
	S_\mu^{{\rm{SH}}}({\vec{f}}) = \frac{1}{1-\mu}\Biggl[ \ln\frac{(1+c)^\mu+(1-c)^\mu}{2^\mu} \Biggr], \\
	S_1^{{\rm{SH}}}({\vec{f}}) = - \frac{1+c}{2}\ln \frac{1+c}{2} - \frac{1-c}{2}\ln \frac{1-c}{2},
\end{eqnarray}
respectively. We find that mode entanglement in the SH state is exponentially sensitive to the displacements $f_k$ (\fref{renyiSH}): If any of the $f_k$ is significantly different from zero the entanglement entropy quickly saturates to its maximum value $\ln 2$.

An effective entanglement Hamiltonian for the reduced qubit state can be constructed such that it has a zero lower energy and an energy gap $\Delta$:
\begin{equation}
	H = \Delta \ket{-}\bra{-},
\end{equation}
where $\ket{-}$ is the Schmidt vector with the (smaller) eigenvalue $\{\frac{1}{2} - \frac{c}{2}\}$. This entanglement Hamiltonian leads to the effective canonical partition function for a two-level system
\begin{equation}
	Z = 1 + \rme^{-\beta\Delta}.
\end{equation}
Defining the effective reciprocal temperature
\begin{equation}\label{shtemp}
	\rme^{-\beta\Delta} = \tanh \vec{f} \cdot \vec{f},
\end{equation}
equates each element of the entanglement spectrum with a corresponding Boltzmann weight of this effective thermal model. Again, correspondence between the single-copy entanglement and $\ln Z$ for a two-level system with an energy gap $\Delta$ follows immediately 
\begin{equation}
	S_\infty^{\rm{SH}}({\vec{f}}) = \ln\left[\frac{2}{1+\exp({-2 {{\rm{\textbf{f}}}}\cdot\vec{f}})}\right] = \ln (1 + {\rme^{-\beta\Delta}}) = \ln Z.
\end{equation}
The effective free energy $F = -S_\infty/\beta$ is 
\begin{equation}
	F = \frac{1}{\beta}\ln\left[\frac{1+\exp({-2 {{\rm{\textbf{f}}}}\cdot\vec{f}})}{2}\right]  = \frac{1}{\beta}\ln \left[\frac{1}{1 + {\rme^{-\beta\Delta}}}\right].
\end{equation}


\section{Concluding Remarks}
We have derived exact expressions for the R\`enyi entanglement entropy in bosonic two-mode squeezed states and the Silbey--Harris polaron ansatz for a qubit interacting with several boson modes. These exact results allowed us to (1) measure the entanglement present due to the interactions between the mode partitions, (2) construct effective canonical distributions with the same probability distributions as the reduced density operators, (3) express the effective reciprocal temperatures in terms of the physical variables such as the squeezing magnitudes and displacement parameters, and (4) identify the single-copy entanglement as the logarithm of an effective partition function.

In particular, we have demonstrated that in a two-mode squeezed vacuum state characterized by the squeezing parameter $z$, the R\'enyi entanglement entropy increases monotonically with $r = |z|$. Also, tracing out the effects of one boson mode leaves the other in a mixed state with identical statistical properties as that of a harmonic oscillator of frequency $\omega$ at reciprocal temperature $\beta = -(\ln\tanh^2 r)/\hbar\omega$. Increasing the squeezing magnitude increases entanglement and the effective temperature of the reduced subsystem. This mode entanglement is not affected by displacing either of the modes before or after squeezing. 

We further determined that entanglement between the qubit and a boson bath in a multi-mode Silbey--Harris state saturates exponentially to its maximum value of $\ln 2$ when any of the boson modes are significantly displaced. In terms of the displacement parameter $\vec{f}$, the reduced qubit subsystem is at an effective reciprocal temperature given by $\beta = -(\ln \tanh \vec{f}\cdot\vec{f})/\Delta$, with $\Delta$ being the energy gap between the qubit states.

\textit{Note added in proof:} Effective thermal theories have also been constructed for entangled left- and right-moving bosonic modes in boundary states of some conformal field theories \cite{zayas2015a}. 

\ack
This work is supported by the University of the Philippines OVPAA through Grant number OVPAA-BPhD-2012-05. KKRS acknowledges support from the DOST Science Education Institute through its Merit Scholarship Program.

\section*{References}

\begin{thebibliography}{10}
\expandafter\ifx\csname url\endcsname\relax
  \def\url#1{{\tt #1}}\fi
\expandafter\ifx\csname urlprefix\endcsname\relax\def\urlprefix{URL }\fi
\providecommand{\eprint}[2][]{\url{#2}}

\bibitem{feynman1963a}
Feynman R~P and Vernon Jr F~L 1963 The theory of a general quantum system
  interacting with a linear dissipative system {\em Ann.~Phys. \rm{(N.~Y.)}\/}
  {\bf 24} 118--73

\bibitem{srednicki1993a}
Srednicki M 1993 Entropy and area {\em Phys. Rev. Lett.\/} {\bf 71} 666--9

\bibitem{wolf2008a}
Wolf M~M, Verstraete F, Hastings M~B and Cirac J~I 2008 Area laws in quantum
  systems: {M}utual information and correlations {\em Phys. Rev. Lett.\/} {\bf
  100} 070502

\bibitem{eisert2010a}
Eisert J, Cramer M and Plenio M~B 2010 \textit{Colloquium}: Area laws for the
  entanglement entropy {\em Rev. Mod. Phys.\/} {\bf 82} 277--306

\bibitem{ryu2006b}
Ryu S and Takayanagi T 2006 Aspects of holographic entanglement entropy {\em J.
  High Energy Phys.\/} {\bf 2006} 045

\bibitem{calabrese2009a}
Calabrese P and Cardy J 2009 Entanglement entropy and conformal field theory
  {\em J.~Phys.~\rm{A}: Math. Theor.\/} {\bf 42} 504005

\bibitem{han1999a}
Han D, Kim Y and Noz M~E 1999 Illustrative example of {F}eynman's rest of the
  universe {\em Am. J. Phys.\/} {\bf 67} 61--6

\bibitem{botero2003a}
Botero A and Reznik B 2003 Modewise entanglement of {G}aussian states {\em
  Phys. Rev. \rm{A}\/} {\bf 67} 052311

\bibitem{silbey1984a}
Silbey R and Harris R~A 1984 Variational calculation of the dynamics of a two
  level system interacting with a bath {\em J.~Chem.~Phys.\/} {\bf 80} 2615--7

\bibitem{vidal1999a}
Vidal G 1999 Entanglement of pure states for a single copy {\em Phys. Rev.
  Lett.\/} {\bf 83} 1046--9

\bibitem{peschel2005}
Peschel I and Zhao J 2005 On single-copy entanglement {\em J. Stat. Mech.\/}
  {\bf 2005} P11002

\bibitem{amico2008a}
Amico L, Fazio R, Osterloh A and Vedral V 2008 Entanglement in many-body
  systems {\em Rev. Mod. Phys.\/} {\bf 80} 517--76

\bibitem{horodecki2009a}
Horodecki R, Horodecki P, Horodecki M and Horodecki K 2009 Quantum entanglement
  {\em Rev. Mod. Phys.\/} {\bf 81} 865--942

\bibitem{peschel2012a}
Peschel I 2012 Special review: {E}ntanglement in solvable many-particle models
  {\em Braz. J. Phys.\/} {\bf 42} 267--91

\bibitem{braunstein2005a}
Braunstein S~L and Van~Loock P 2005 Quantum information with continuous
  variables {\em Rev. Mod. Phys.\/} {\bf 77} 513--77

\bibitem{eichler2011a}
Eichler C, Bozyigit D, Lang C, Baur M, Steffen L, Fink J~M, Filipp S and
  Wallraff A 2011 Observation of two-mode squeezing in the microwave frequency
  domain {\em Phys. Rev. Lett.\/} {\bf 107} 113601

\bibitem{parkins2006a}
Parkins A~S, Solano E and Cirac J~I 2006 Unconditional two-mode squeezing of
  separated atomic ensembles {\em Phys. Rev. Lett.\/} {\bf 96} 053602

\bibitem{esteve2008a}
Est\`eve J, Gross C, Weller A, Giovanazzi S and Oberthaler M~K 2008 Squeezing
  and entanglement in a {B}ose--{E}instein condensate {\em Nature\/} {\bf 455}
  1216--9

\bibitem{xue2007a}
Xue F, Liu Y~X, Sun C~P and Nori F 2007 Two-mode squeezed states and entangled
  states of two mechanical resonators {\em Phys. Rev. B\/} {\bf 76} 064305

\bibitem{vidal2000a}
Vidal G 2000 Entanglement monotones {\em J. Mod. Opt.\/} {\bf 47} 355--76

\bibitem{li2008a}
Li H and Haldane F 2008 Entanglement spectrum as a generalization of
  entanglement entropy: Identification of topological order in non-{A}belian
  fractional quantum {H}all effect states {\em Phys. Rev. Lett.\/} {\bf 101}
  010504

\bibitem{nielsenchuang}
Nielsen M~A and Chuang I~L 2010 {\em Quantum Computation and Quantum
  Information\/} (Cambridge: Cambridge University Press)

\bibitem{peschel2011a}
Peschel I and Chung M~C 2011 On the relation between entanglement and subsystem
  {H}amiltonians {\em Europhys. Lett.\/} {\bf 96} 50006

\bibitem{dechiara2012a}
De~Chiara G, Lepori L, Lewenstein M and Sanpera A 2012 Entanglement spectrum,
  critical exponents, and order parameters in quantum spin chains {\em Phys.
  Rev. Lett.\/} {\bf 109} 237208

\bibitem{wong2013a}
Wong G, Klich I, Pando Zayas L A and Vaman D 2013 Entanglement temperature and
  entanglement entropy of excited states {\em J. High Energy Phys.\/} {\bf
  2013} 20

\bibitem{chandran2014a}
Chandran A, Khemani V and Sondhi S~L 2014 How universal is the entanglement
  spectrum? {\em Phys. Rev. Lett.\/} {\bf 113} 060501

\bibitem{schliemann2014a}
Schliemann J 2014 Entanglement thermodynamics {\em J. Stat. Mech.\/} {\bf 2014}
  P09011

\bibitem{adesso2005a}
Adesso G, Serafini A and Illuminati F 2005 Entanglement, purity, and
  information entropies in continuous variable systems {\em Open Syst. Inf.
  Dyn.\/} {\bf 12} 189--205

\bibitem{adesso2012a}
Adesso G, Girolami D and Serafini A 2012 Measuring {G}aussian quantum
  information and correlations using the {R}\'enyi entropy of order 2 {\em
  Phys. Rev. Lett.\/} {\bf 109} 190502

\bibitem{jonathan1999a}
Jonathan D and Plenio M~B 1999 Minimal conditions for local pure-state
  entanglement manipulation {\em Phys. Rev. Lett.\/} {\bf 83} 1455--8

\bibitem{eisert2005a}
Eisert J and Cramer M 2005 Single-copy entanglement in critical quantum spin
  chains {\em Phys.~Rev.~{\rm{A}}\/} {\bf 72} 042112

\bibitem{hadley2008a}
Hadley C 2008 Single-copy entanglement in a gapped quantum spin chain {\em
  Phys.~Rev.~Lett.\/} {\bf 100} 177202

\bibitem{loudon1987a}
Loudon R and Knight P~L 1987 Squeezed light {\em J. Mod. Opt.\/} {\bf 34}
  709--59

\bibitem{ekert1989a}
Ekert A~K and Knight P~L 1989 Correlations and squeezing of two-mode
  oscillations {\em Am. J. Phys.\/} {\bf 57} 692--7

\bibitem{hongyi1996a}
Hong-yi F and Yue F 1996 Representations of two-mode squeezing transformations
  {\em Phys. Rev. \rm{A}\/} {\bf 54} 958--60

\bibitem{kurochkin2014a}
Kurochkin Y, Prasad A~S and Lvovsky A~I 2014 Distillation of the two-mode
  squeezed state {\em Phys. Rev. Lett.\/} {\bf 112} 070402

\bibitem{chin2011a}
Chin A~W, Prior J, Huelga S~F and Plenio M~B 2011 Generalized polaron ansatz
  for the ground state of the sub-ohmic spin-boson model: {A}n analytic theory
  of the localization transition {\em Phys. Rev. Lett.\/} {\bf 107} 160601

\bibitem{zayas2015a}
Pando Zayas, L~A and Quiroz, N 2015 Left-right entanglement entropy of boundary states {\em JHEP\/} {\bf 2015} 110


\end{thebibliography}

\providecommand{\newblock}{}

\end{document}